# Ferroelectricity controlled chiral spin textures and anomalous valley Hall effect in the Janus magnet-based multiferroic heterostructure

Yingmei Zhu[1], Qirui Cui[1,3], Jinghua Liang[1], Yonglong Ga[1], Hongxin Yang[1,2,*]

[1] *Ningbo Institute of Materials Technology and Engineering, Chinese Academy of Sciences, Ningbo 315201, China*

[2] *Center of Materials Science and Optoelectronics Engineering, University of Chinese Academy of Sciences, Beijing 100049, China*

[3] *Faculty of Science and Engineering, University of Nottingham Ningbo China, Ningbo 315100, China*

[*]Email: hongxin.yang@nimte.ac.cn

# Abstract

Realizing effective manipulation and explicit identification of topological spin textures are two crucial ingredients to make them as information carrier in spintronic devices with high storage density, high data handling speed and low energy consumption. Electric-field manipulation of magnetism has been achieved as a dissipationless method compared with traditional regulations. However, the magnetization is normally insensitive to the electric field since it does not break time-reversal symmetry directly, and distribution of topological magnetic quasiparticles is difficult to maintain due to the drift arising from external fluctuation, which could result in ambiguous recognition between quasiparticles and uniform magnetic background. Here, we demonstrate that electric polarization-driven skyrmionic and uniform ferromagnetic states can be easily and explicitly distinguished by transverse voltage arising from anomalous valley Hall effect in the Janus magnet-based multiferroic heterostructure LaClBr/In$_2$Se$_3$. Our work provides an alternative approach for data encoding, in which data are encoded by combing topological spin textures with detectable electronic transport.

# 1. Introduction

Controlling magnetism by electric field is the fundamental avenue for realizing high-efficiency and extremely low-power consumption memories and logic devices [1-5], such as electric field-tunable magnetoelectric random access memories (MERAMs) with low writing energy [6]. Multiferroic materials, with both ferromagnetism and ferroelectricity, are prospective candidates that can achieve electric-field modulation of ferromagnetism, magnetic direction, and magnetic anisotropy via the magnetoelectric coupling effect [7-10]. The topological magnetic quasiparticles, consisting of skyrmion and bimeron, have attracted intensive attention. Due to possess many advantages such as nonvolatility, small-size, and high mobility, these quasiparticles hold huge potential in spintronic devices for information technologies [11-13]. In 2016, it was reported that the skyrmion can be transformed to uniform ferromagnetism by electric field Fe/Ir(111) film with a scanning tunneling microscope (STM) [14]. Then, ferroelectrically tunable magnetic skyrmions have been demonstrated in ultrathin oxide $BaTiO_3$/$SrRuO_3$ heterostructure (HS) [15]. Compared with bulk or multilayered multiferroic materials, the two-dimensional (2D) van der Waals (vdW) magnetic/ferroelectric HSs promote much miniaturization, simpler structure, better tunability, and highly interface quality for spintronic devices [16]. Recently, writing and erasing of skyrmion and bimeron also has achieved in 2D multiferroic HSs $MnBi_2Se_2Te_2$/$In_2Se_3$, $LaCl$/$In_2Se_3$, $WTe_2$/$CrCl_3$/$CuInP_2S_6$, and $Fe_3GeTe_2$/$In_2Se_3$, where $MnBi_2Se_2Te_2$, $LaCl$, $WTe_2$/$CrCl_3$ and $Fe_3GeTe_2$ are vdW magnets with Dzyaloshinskii-Moriya interaction (DMI), while $In_2Se_3$ is an atomic FE material easily altered by out-of-plane (OOP) electric field [17-20]. The topological magnetic quasiparticles and ferromagnetic (FM) states are potentially encoded and stored as '1' and '0' bit carriers, respectively. However, the location of these quasiparticles is susceptible to external fluctuations, making it difficult to distinguish from FM state. This problem has been mentioned in the skyrmion-based racetrack-type memories, which need to be maintained skyrmion/bimeron distribution in the operational time scales [21, 22]. Notably, the reversal of FE polarization induces different electronic states of 2D FE-based vdW HSs in reciprocal space. For example, the switchable topological edge states, valley polarization, and metal-to-semiconductor phase transition are found in bilayer-Bi(111)/$In_2Se_3$ and $β$-Sb/$In_2Se_3$,

HfN$_2$/CrI$_3$/In$_2$Se$_3$, and CrI$_3$/Sc$_2$CO$_2$, respectively [23-26]. The band structures of a large variety of FE-aided vdW HS can be obviously regulated because the two surface of FE layer is nonequivalent induced by built-in electric field [27]. The above results imply that the variation of band structures controlled by FE polarization is hopefully used in unambiguously distinguishing spin configurations.

Here, employing the first-principles calculations and atomistic spin model simulations, we propose a type of magnetoelectric effect which can be realized in 2D Janus-based multiferroic vdW HS, La$XY$/In$_2$Se$_3$ ($X/Y$=Cl, Br, I, $X\neq Y$), where La$XY$ can be constructed by replacing one of halogen atom in 2$H$-La$X_2$ and considered as an ideal candidate to realize FE-controlled electronic states and topological magnetic quasiparticles since it combines long-range ferromagnetism, robust valley polarization and inversion symmetry breaking-allowed DMI [28]. Moreover, La$XY$ monolayer has small lattice mismatch with room-temperature ferroelectric materials In$_2$Se$_3$. For LaClBr/In$_2$Se$_3$ HS, skyrmions embedded in domain walls with metal state are transformed into uniform ferromagnetism with anomalous valley Hall effect (AVHE) by switching the direction of FE polarization from -$z$ (P↓) to +$z$ (P↑) axis, implying the skyrmionic and uniform FM state can be distinguished by transverse voltage of sample. Furthermore, skyrmion (bimeron) solitons in La$XY$/P↓ are obtained by applying external magnetic field. Further analysis indicates that the topological magnetic phase transition attributed to the change of DMI caused by FE polarization, and the polarization-dependent band alignments are influenced by charge transfer between layers. Our work thus propose that approach for realizing unambiguous recognition of FE-controlled topological spin configurations under the assistance of band structures, and provides specific systems for realizing it.

## 2. Calculation method

All first-principles calculations are performed using the density functional theory (DFT) as implemented in the Vienna ab initio Simulation Package (VASP) code [29-31]. The exchange and correlation functionals are treated by the generalized gradient approximation (GGA) of the Perdew-Burke-Ernzerhof functional (PBE) [32, 33]. The interaction between ions and electrons is described by projector-augmented wave (PAW) method [34]. After compared the magnetic parameters under the different cut-off energies (see Table S1 and Section S4 in Supplementary material), the plane-wave basis set in 420 eV. To avoid interaction between adjacent layers, the thickness of vacuum layer is set no less than 15 Å along the *z* direction. We employ the GGA+*U* method with *U* = 7 eV to treat the *f* electrons for La as reported in the previous studies [18]. Brillouin zone is sampled using Γ-centered 24×24×1 Monkhorst-pack *k*-point mesh. The electronic convergence is performed with a tolerance of $10^{-7}$ eV. Optimized structures are fully relaxed until the force converged on each atom less than $10^{-3}$ eV/Å. To check the stability of the Janus La*XY* monolayer, Phonon dispersions are calculated with 4×4×1 and 5×5×1 supercell by using the PHONOPY code [35], and *ab initio* molecular dynamics (AIMD) are simulated with 4×4×1 supercell in the canonical *NVT* ensemble [36]. The vdW interaction between La*XY* and $In_2Se_3$ monolayer are corrected by DFT-D3 [37]. To avoid the influence of external strain from sublattice [Table S2 and Section S3 of supplementary material], by using 1×1 cell of $In_2Se_3$ to match 1×1 cell of La*XY*, the lattice mismatch is 0.51%, 1.85%, and 3.09% for LaClBr/P, LaClI/P, and LaBrI/P, respectively. The maximally location Wannier function are calculated using WANNIER90 package code to obtain the Berry curvature and anomalous Hall conductivity [38]. More computational details are demonstrated in Supplementary material.

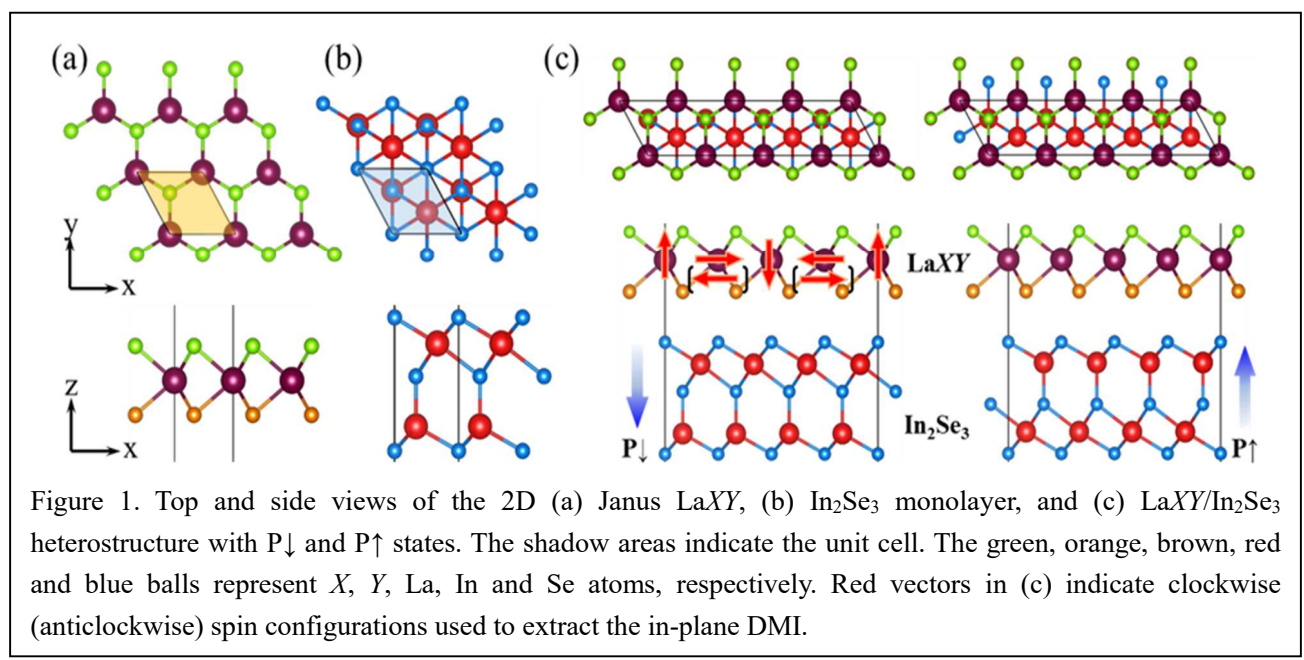

Figure 1. Top and side views of the 2D (a) Janus La*XY*, (b) $In_2Se_3$ monolayer, and (c) La*XY*/$In_2Se_3$ heterostructure with P↓ and P↑ states. The shadow areas indicate the unit cell. The green, orange, brown, red and blue balls represent *X*, *Y*, La, In and Se atoms, respectively. Red vectors in (c) indicate clockwise (anticlockwise) spin configurations used to extract the in-plane DMI.

## 3. Results and discussions

Figure 1(a) displays the top and side views of Janus La$XY$ ($X/Y$=Cl, Br, I, $X \neq Y$), which contain two atomic planes with different halogen elements and form hexagonal network with point group $C_{3v}$. Bulk LaBr$_2$ is a hexagonal layered crystal with a space group of P6$_3$/mmc [39], and LaBr$_2$ monolayer has been proposed as an ideal ferromagnetic semiconductor with Tc>200K [40,41], Which is higher than that of CrI$_3$ monolayer (45K) [42] and Cr$_2$Ge$_2$Te$_6$ bilayer (28K) [43]. The optimized lattice constant of LaClBr, LaClI, and LaBrI monolayer is listed in Table 1, and corresponding magnetic moment of La atom is 0.385, 0.374, and 0.371 μ$_B$, respectively. The structural stability, including the phonon dispersions (Figure S1 shows that the other two monolayers are dynamically stable except for LaClI with small imaginary frequency around Γ point) and molecular dynamic simulations (Figure S2 shows that La$XY$ monolayers are thermal stability), are demonstrated in Section S1 and S2 of Supplementary material. In$_2$Se$_3$ is a 2D vdW room-temperature FE material [44-46] and Figure 1(b) presents its crystal structure, which the lattice constant is 4.108 Å. The position of center-Se layer in In$_2$Se$_3$ gives rise to the different OOP polarization P↓ and P↑, leading to the variation of magnetic and electronic properties in La$XY$ when In$_2$Se$_3$ is used as substrate. The successful synthesis of Janus monolayers of transition metal dichalcogenides (MoSSe, MoSH) and FE-aided HSs (MoS$_2$/In$_2$Se$_3$) illustrate the experimental feasibility of construction of La*XY*/In$_2$Se$_3$ multiferroic HSs [47-50]. The La*XY* monolayers perform in-plane anisotropy, ferromagnetic state and clockwise chirality DMI [see Table S2]. As described in Section S3 of Supplementary material, we consider twelve stacking configurations of La*XY*/P↓ [Figure S3] and calculate the corresponding total energy [Figure S4]. We find that the P↓ state is always energetically more stable than P↑ state, which is consistent with previously generic trend [27]. The magnetic properties can be regulated by changing the interfacial coupling, when revers the OOP polarization of In$_2$Se$_3$ in the La*XY*/In$_2$Se$_3$ HS. In Table 1, the equilibrium interlayer distance of La*XY*/P↓ is lower than P↑ due to the stronger interfacial coupling between La*XY* and In$_2$Se$_3$ in P↑. Top and side views of the most stable structure of LaClBr/P↓ and LaClBr/P↑ are shown in Figure 1(c), where the electric polarization is -0.103 and 0.113 eÅ/u.c., respectively. The polarization of LaClBr/In$_2$Se$_3$ is very close to that of In$_2$Se$_3$ monolayer (0.1 eÅ/u.c.), indicating that the ferroelectricity of HS mainly comes from In$_2$Se$_3$ layer. In experiment, an OOP electrical field around 1 V/nm can flip the perpendicular polarization direction [51], which implies that the reversal of polar orientation is achieved using a large enough electric field in La*XY*/In$_2$Se$_3$ HS.

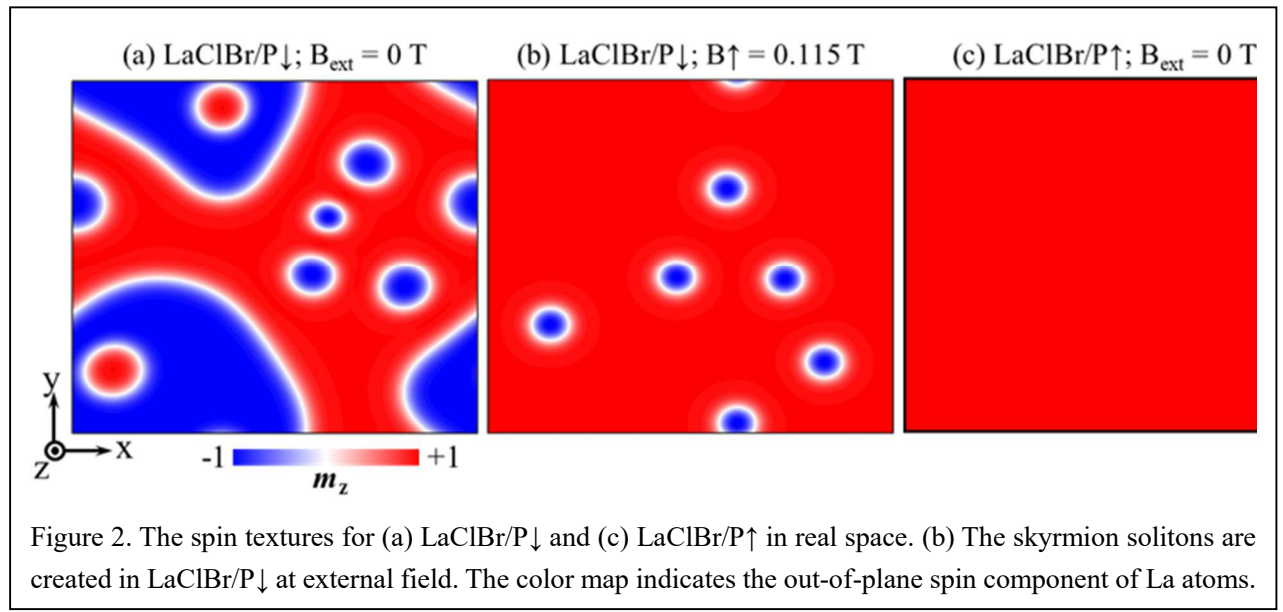

Figure 2. The spin textures for (a) LaClBr/P↓ and (c) LaClBr/P↑ in real space. (b) The skyrmion solitons are created in LaClBr/P↓ at external field. The color map indicates the out-of-plane spin component of La atoms.

Table 1. Calculated lattice constants $a$ (Å), magnetic moment of La atom $m_{La}$ ($\mu_B$), equilibrium interlayer distance $d$ (Å), magnetic anisotropy energy $K$ (meV), nearest-neighbor exchange coupling $J$ (meV), and in-plane DMI component $d_\parallel$ (meV) of La*XY*/P↓ and La*XY*/P↑ (*X*, *Y* = Cl, Br, I).

|  | $a$ | $m_{La}$ | $d$ | $K$ | $J_{xx}$ | $J_{yy}$ | $J_{zz}$ | $J$ | $d_\parallel$ |
|---|---|---|---|---|---|---|---|---|---|
| LaClBr/P↓ | 4.087 | 0.368 | 3.073 | 0.010 | 7.809 | 7.803 | 7.788 | 7.800 | 0.197 |
| LaClBr/P↑ | 4.087 | 0.381 | 3.101 | 0.009 | 6.757 | 6.754 | 6.739 | 6.750 | -0.008 |
| LaClI/P↓ | 4.184 | 0.351 | 3.172 | -0.050 | 8.141 | 8.124 | 8.073 | 8.113 | 0.597 |
| LaClI/P↑ | 4.184 | 0.364 | 3.224 | -0.090 | 7.071 | 7.069 | 7.013 | 7.051 | 0.167 |
| LaBrI/P↓ | 4.235 | 0.343 | 3.136 | -0.011 | 8.280 | 8.271 | 8.239 | 8.263 | 0.424 |
| LaBrI/P↑ | 4.235 | 0.358 | 3.195 | -0.073 | 7.281 | 7.277 | 7.222 | 7.260 | 0.158 |

In order to investigate the magnetic properties of La*XY*/In$_2$Se$_3$, we adopt the following spin Hamiltonian:

$$H = -K \sum_i (S_i^z)^2 - J \sum_{\langle i,j \rangle} \mathbf{S}_i \cdot \mathbf{S}_j - \sum_{\langle i,j \rangle} \mathbf{D}_{ij} \cdot (\mathbf{S}_i \times \mathbf{S}_j) - \mu_{La} B \sum_i S_i^z \quad (1)$$

where $S_i$ is a unit vector representing the orientation of the spin of the $i$th La atom, and $\langle i,j \rangle$ represents the nearest-neighbor (NN) La atom pairs. The magnetic parameters $K$, $J$, and $D_{ij}$ in the first three terms represent the magnetic anisotropy, exchange coupling, and interatomic DMI, respectively. And the $\mu_{La}$ and $B$ in the last term represent the magnetic moment of La and external magnetic field, respectively. Here we adopt the sign convention that $K>0$ and $K<0$ represent perpendicular magnetic anisotropy (PMA) and in-plane magnetic anisotropy (IMA), $J>0$ and $J<0$ indicates the FM and antiferromagnetic coupling, and the in-plane (IP) DMI component $d_\parallel>0$ and $d_\parallel<0$ favor spin configuration with anticlockwise (ACW) and clockwise

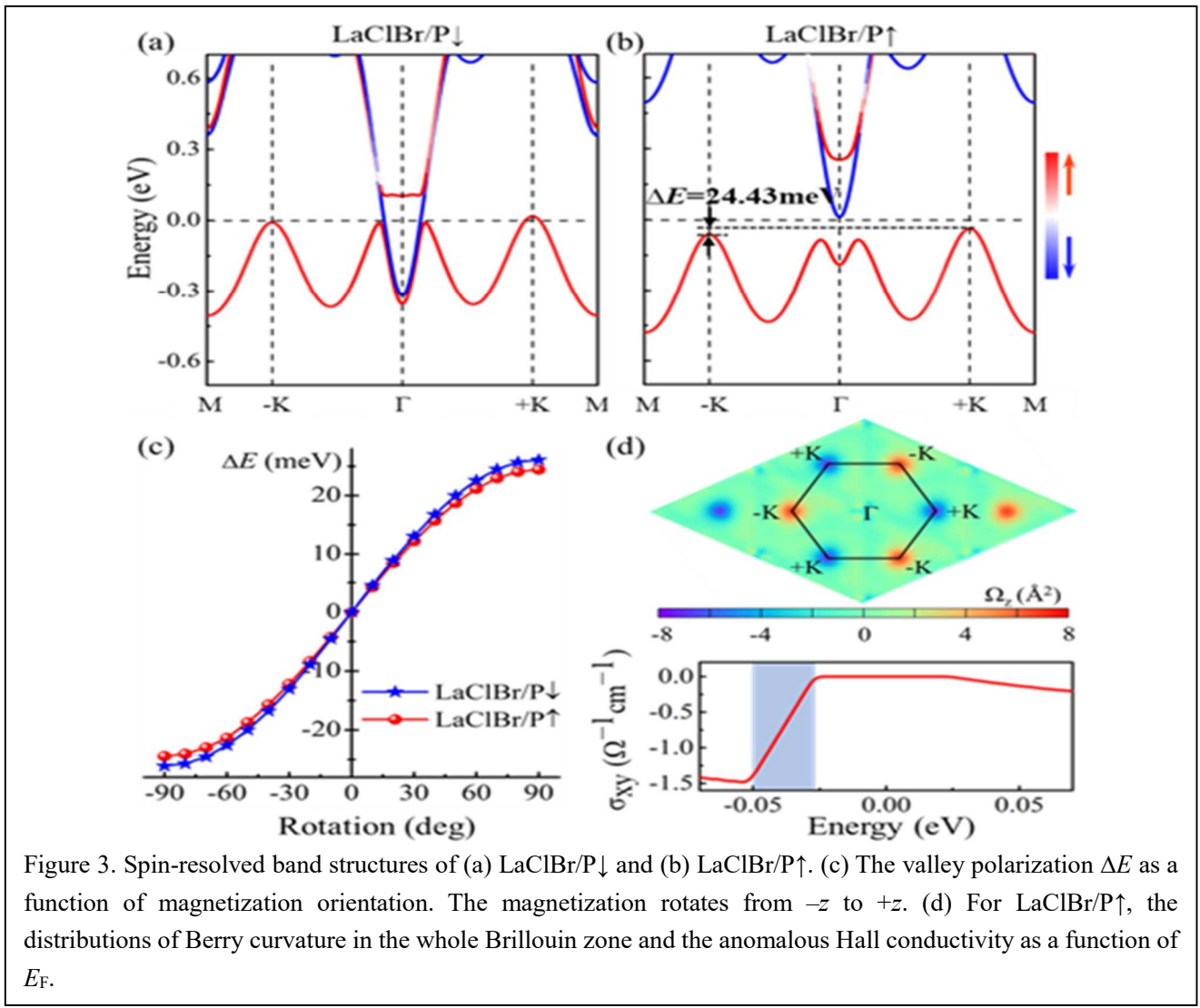

Figure 3. Spin-resolved band structures of (a) LaClBr/P↓ and (b) LaClBr/P↑. (c) The valley polarization $\Delta E$ as a function of magnetization orientation. The magnetization rotates from $-z$ to $+z$. (d) For LaClBr/P↑, the distributions of Berry curvature in the whole Brillouin zone and the anomalous Hall conductivity as a function of $E_F$.

(CW) [as marked in Figure 1(c) with red vectors], respectively. We first calculate the magnetic parameters of LaClBr/In$_2$Se$_3$ with Br-Se and Cl-Se interface contact, which constructed by switching LaClBr layer along $z$ direction of the most stable structure [Table S3 and Section S3 of Supplementary material]. When the interface contacts between LaClBr and In$_2$Se$_3$ changes, i.e. LaClBr layer is switched along $z$ direction, DMI chirality is consequently reversed according to Maryia rules [52]. That indicates the interface contact has a great influence on these systems, then we use the most stable structure in the following calculations [Figure S4]. The calculated details about these parameters are explained in Section S4 of Supplementary material, and the results are shown in Table 1. The magnetic moment of La atom of La$XY$/In$_2$Se$_3$ is slightly smaller than that of pristine La$XY$. Compared with pristine La$XY$, La$XY$/P↓(↑) except for LaClBr/P↑ HS possess weaker in-plane anisotropy, higher exchange coupling and larger DM interaction. The magnetic anisotropy energy (MAE) shows that LaClBr/In$_2$Se$_3$ has weak PMA whose amplitude reaches to 0.010 and 0.009 meV for P↓ and P↑, respectively. The LaClI/In$_2$Se$_3$ and LaBrI/In$_2$Se$_3$ both have an easy magnetization plane, i.e., they exhibit IMA. We then calculate the NN exchange coupling with magnetization along positive directions of $x, y, z$ axis, $J_{xx}, J_{yy}, J_{zz}$, and spin-orbit coupling (SOC) effects are also considered in calculations. We find that the NN exchange coupling of all systems remain FM state and are almost isotropic along different directions, indicating that the average of magnetic coupling parameter $J$ can be defined to represents coefficient of the second term in the Eq. (1). The second-nearest- and third-nearest-neighboring exchange couplings are much smaller than $J$ and thus can be neglected in these systems [see Section S4 and Figure S6 of Supplementary material]. For P↓ and P↑, the exchange coupling barely changes. For example, $J$ is tuned from 7.80 to 6.75 meV in LaClBr/In$_2$Se$_3$ as the electric polarization reverses. Therefore, the electric polarization does not directly affect magnetic ordering of La$XY$ layer. In order to determine the transition temperature of the LaClBr/P↓(↑), we simulate the Curie temperature ($T_c$) using Monte Carlo method. As shown in Figure S7, the Curie temperatures of La$XY$/P↓ are around 100K, and those of La$XY$/P↑ are around 85K. For In$_2$Se$_3$, the ferroelectricity can sustain over room temperature which has been demonstrated in experimentally [46]. Remarkably, when direction of FE polarization reverses from down to up, the chirality of DMI of LaClBr/In$_2$Se$_3$ changes from ACW to CW, and the amplitude of DMI changes from 0.197 to 0.008 meV. Large variation of amplitude of DMI also emerges in LaClI/In$_2$Se$_3$ and LaBrI/In$_2$Se$_3$, while the chirality of DMI remains ACW. Ferroelectricity controllable DMI is expected to give rise the obvious variation of topological magnetic phases [see later discussions].

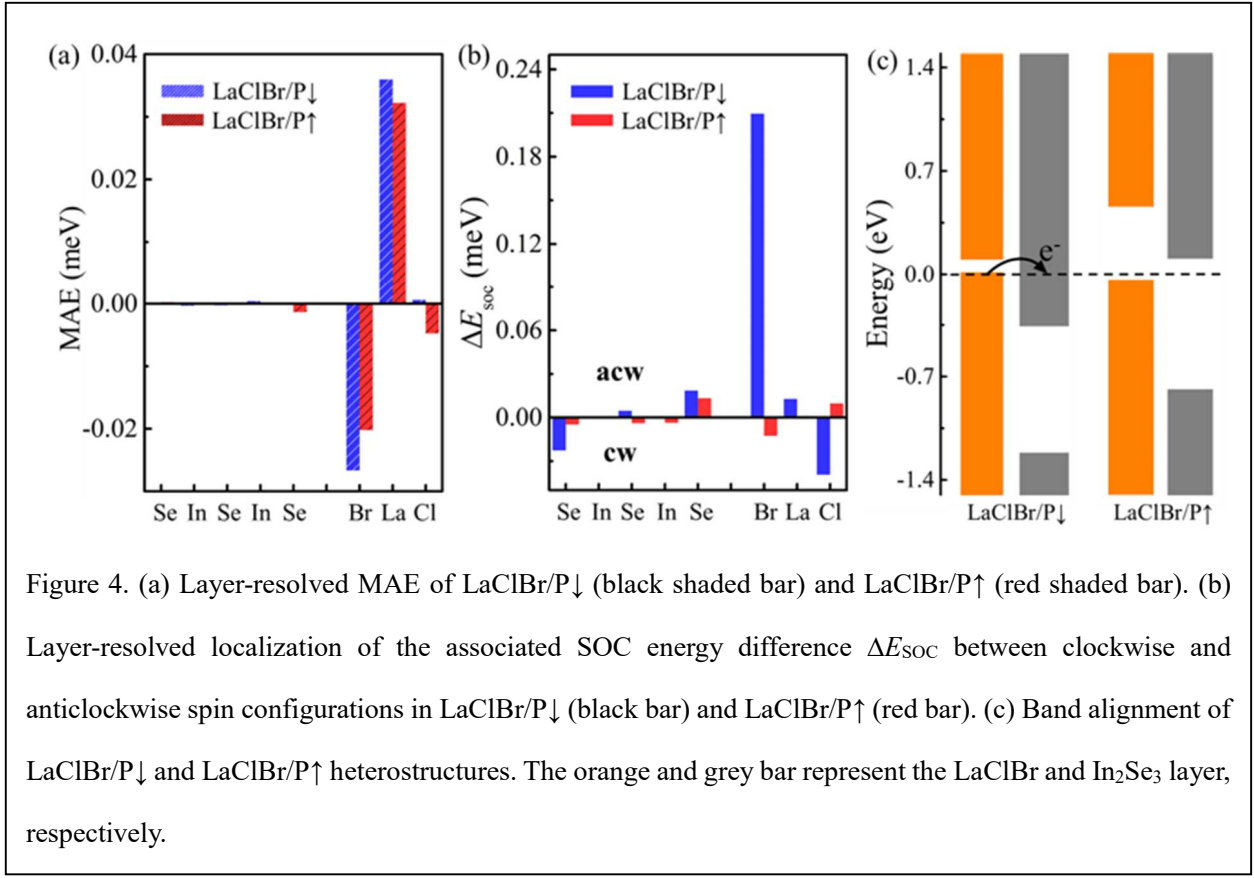

Figure 4. (a) Layer-resolved MAE of LaClBr/P↓ (black shaded bar) and LaClBr/P↑ (red shaded bar). (b) Layer-resolved localization of the associated SOC energy difference $\Delta E_{SOC}$ between clockwise and anticlockwise spin configurations in LaClBr/P↓ (black bar) and LaClBr/P↑ (red bar). (c) Band alignment of LaClBr/P↓ and LaClBr/P↑ heterostructures. The orange and grey bar represent the LaClBr and In$_2$Se$_3$ layer, respectively.

Once all the magnetic parameters $J$, $K$, and $d_\parallel$ in the spin Hamiltonian are calculated by first-principles calculations, we can perform the atomistic spin model simulations of La$XY$/In$_2$Se$_3$ HS using the VAMPIRE package in a large enough space with a square length of 300 nm [53]. The initial spin direction is set to random states in the periodic boundary. Section S5 in Supplementary material introduces more simulation details. The topological charge, $Q = \frac{1}{4\pi} \int \boldsymbol{S} \cdot (\partial_x \boldsymbol{S} \times \partial_y \boldsymbol{S}) dxdy$, is adopted to quantify information of the skyrmion or bimeron, which all can be ascribed to $Q = \pm 1$. In the LaClBr/In$_2$Se$_3$ HS, skyrmions embedded in domain walls transform into uniform OOP FM states are achieved by reversing the polarization from P↓ to P↑, as shown in Figures 2(a) and (c), which origin from the significant difference of DMI strengths. Furthermore, in Figure 2(b), we find the skyrmion solitons with $Q = -1$ is generated in LaClBr/P↓ by applying an OOP magnetic field along with the positive direction of the $z$ axis (B↑). For P↓ state of LaClI/In$_2$Se$_3$ and LaBrI/In$_2$Se$_3$ HS, we obtain the chiral Néel domain wall (DW) arising from strong DMI [see Figures S8(a) and (e)]. It is worth noting that the DMI/exchange coupling ratio ($|d_\parallel/J|$) in LaClI/P↓ (0.074) is larger than in LaBrI/P↓ (0.051), which is consistent with the smaller width of the domain and DW energy in LaClI/P↓ compared with LaBrI/P↓ HS. Especially, isolated skyrmions are induced by using B↑ in LaClI (LaBrI)/P↓ [Figures S8(b) and (f)], and bimerons consisted with a vortex and an anti-vortex appear in LaClI (LaBrI)/P↓ by applying an IP magnetic field along with +x axis (B→) [Figures S8(c) and (g)]. When the polarization is reversed to the P↑ state, the reduced DMI leads to small $|d_\parallel/J|$ of 0.024 and 0.022 for LaClI/P↑ and LaBrI/P↑, respectively, accompanied by IP FM state [Figures S8(d) and (h)]. The chiral spin textures and FM states are used to '1' and '0' bit carriers by switching the FE polarization for encoding and storing information. However, the location of these spin textures is susceptible to thermal fluctuations or electric current, making it difficult to distinguish from FM state.

Next, we focus on the electronic properties of La$XY$/In$_2$Se$_3$ HS and use it to realize unambiguously recognition of different signal. Figures 3(a) and (b) show the spin-resolved band structures with SOC of LaClBr/In$_2$Se$_3$ HS, where the conduction bands and valence bands near the Fermi level consisting of carriers with the opposite spin. The LaClBr/P↓ performs metallic character while LaClBr/P↑ possesses an indirect band gap of 48 meV, and the nondegenerate valleys appear in the VBM at the -K and +K points. We also find that the metal state for LaClBr/P↓ preserves when the magnetic orientation rotates, as shown in Figure S9. The easy magnetization orientation of LaClBr/In$_2$Se$_3$ is along +z and it can be turned form OOP to IP by overcoming an energy barrier of 10 and 9 meV for LaClBr/P↓ and P↑, respectively. When we rotate the magnetization from -z to +z axis, the valley polarization for P↓(P↑) defined as the energy different between VBM at +K and -K, $\Delta E = E_{VBM,+K} - E_{VBM,-K}$, varies from -26.09 (-24.43) to 26.09 (24.43) meV [Figure 3(c)]. The quantitative relationship between magnetization orientation and valley polarization is consistent with previous conclusion, i.e., $\Delta E \propto 4\cos\theta$, where the $\theta$ is the angle between the direction of spin and +z axis [17]. That means the valley splitting will gradually disappear as spin rotates from OOP to IP, which has been confirmed by first-principles calculations in Figure 3(c). To demonstrate the AVHE which arises from the valley polarization, we calculate the Berry curvature of LaClBr/P↑ HS according to the Kubo formula as [54]:

$$\Omega(k) = -\sum_n \sum_{n \neq n'} f_n \frac{2Im\langle\psi_{nk}|v_x|\psi_{n'k}\rangle\langle\psi_{n'k}|v_y|\psi_{nk}\rangle}{(E_n-E_{n'})^2} \quad (2)$$

where $f_n$ is Fermi-Dirac distribution function, $v_{x(y)}$ is velocity operator along $x(y)$ direction, and $\psi_{nk}$ is Bloch wave function with eigenvalue $E_n$. Figure 3(d) shows the Berry curvatures along the high-symmetry point in the whole 2D Brillouin zone, and the opposite signs and different absolute values are acquired at the -K and +K points while the values close to zero at other points. The anomalous valley Hall conductivity $\sigma_{xy}$ can be obtained by integrating the Berry curvature, which is expressed as [55]:

$$\sigma_{xy} = \frac{e^2}{h}\frac{1}{2\pi}\int_{BZ} d\mathbf{k}^2 \Omega(\mathbf{k}) \quad (3)$$

As shown by labelled shadow in Figure 3(d), when the $E_F$ is shifted between -K and +K valleys in the valence bands, a nonzero valley-polarized $\sigma_{xy}$ is achieved, proving the existence of AVHE. However, for LaClI/In$_2$Se$_3$ and LaBrI/In$_2$Se$_3$, all band structures maintain metallic character under the influence of FE polarization even if the CBM lifts upward at P↑ state [Figure S10]. Generally speaking, we obtain the topological spin textures with metal state in P↓ and the trivial topological with valley polarization in P↑. Therefore, skyrmion state and uniform ferromagnetic state with transverse voltage arising from AVHE can be used '1' and '0' bit carriers in data storage, which is modulated by switching the FE polarization in LaClBr/In$_2$Se$_3$ HS.

Then we analysis the physical mechanism of the FE-controlled magnetic phase and band structures transition of La*XY*/In$_2$Se$_3$ HS. From the layer-resolved MAE of LaClBr/In$_2$Se$_3$ [Figure 4(a)], LaClI/In$_2$Se$_3$ and LaBrI/In$_2$Se$_3$ [Figure S11(a)], one can see that the MAE mainly comes from La*XY* layer and the top-layer in La*XY* basically does not contribute to the MAE. For LaClBr/In$_2$Se$_3$, the PMA of La is slightly larger than IMA of Br, while for LaClI(LaBrI)/In$_2$Se$_3$, both the La and *X* atoms contribute the IMA and the strength is enhanced significantly in P↑ compared with P↓. As shown in Table S4, the DMI is reduced when the polarization of In$_2$Se$_3$ is changed to zero by placing all atoms at the local centrosymmetric positions, which means the electric field produced by polarization rather than the interfacial hybridization has the great importance in determining DMI. We further elucidate the layer-resolved SOC energy difference Δ$E_{SOC}$ between the CW and ACW configurations for LaClBr/In$_2$Se$_3$ [Figure 4(b)], LaClI/In$_2$Se$_3$ and LaBrI/In$_2$Se$_3$ [Figure S11(b)]. Strikingly, the sign and strength of DMI are mainly caused by the SOC from heavy elements, and the DMI of halogen atom which is close to the In$_2$Se$_3$ layer is increased obviously while the polarization of In$_2$Se$_3$ reverses from P↑ to P↓. The metal-to-semiconductor transition in LaClBr/In$_2$Se$_3$ HS based FE polarization is mainly due to the change of the band edge position of the In$_2$Se$_3$ layer. Figure 4(c) shows the band alignment of the LaClBr/P↓ and P↑ In the P↓ state, the CBM of the In$_2$Se$_3$ layer becomes lower than the VBM of the LaClBr layer, exhibiting a type III band alignment with the vanishing of the band gap, whereas in the case of P↑ state, the system becomes type II semiconductor with the conduction band minimum and valence band maximum derived from the In$_2$Se$_3$ and LaClBr layer, respectively. The fact that FE reversal-induced phase transition gives rise to the shift of electronic states of LaClBr/In$_2$Se$_3$ HS. Bader analysis exhibits that the charge transfer from LaClBr to In$_2$Se$_3$ decreases from 0.035 to 0.004e when FE polarization is changed from down to up [56]. We further calculate magnetic parameters of $K$, $J$, and $d_\parallel$ for LaClBr monolayer under 0.035e hole doping [Table S2]. These parameters are very close to that of LaClBr/P↓ HS, which indicate that the versatile properties in La*XY*/In$_2$Se$_3$ is mainly derived from charge transfer between interface. Moreover, the planer-average charge-density difference shows that the charge is mainly transferred in In$_2$Se$_3$ layer of LaClBr/P↓ but restricted in the interface of LaClBr/P↑ [Figure S12].

## 4. Conclusion

In summary, we systematically investigate the chiral spin configurations in real space and electronic band structures in reciprocal space by modulating FE polarization of 2D vdW multiferroic La*XY*/In$_2$Se$_3$ HS via the first-principles calculations and atomistic simulation. The strength of DMI decrease apparently when the polar orientation is turned from P↓ to P↑, resulting in skyrmions in domain walls are transformed into uniform ferromagnetism. By applying OOP or IP magnetic field, the skyrmion or bimeron solitons can be achieved in P↓, respectively. The generation of these topological magnetic quasiparticles could be used as a carrier of information transmission in spintronic devices. On the other hand, the electronic band structures are modulated easily by reversal of electrical polarization of La*XY*/In$_2$Se$_3$ HS.

Especially for LaClBr/In$_2$Se$_3$ HS, the metal-to-semiconductor phase transition is realized by flipping the polarization states from down to up, accompanied with valley polarization emergence. Detailed analysis demonstrates that charge transfers between LaClBr and In$_2$Se$_3$ layer play a crucial role in band alignments of HS: when P↓ is reversed to P↑ state, the conduction bands are lifted away from the Fermi level due to the reduce of the interlayer transferred charge. Therefore, the unambiguous recognition of different spin configurations is realized under the assistance of AVHE. Our work provides an alternative approach for data encoding, in which a data can be encoded by combing chiral spin configuration with untrivial electronic effects.

## Acknowledgments

This work was supported by "Pioneer" and "Leading Goose" R&D Program of Zhejiang Province under Grant No. 2022C01053, the Key Research Program of Frontier Sciences, CAS (Grant NO. ZDBS-LY-7021); National Natural Science Foundation of China (Grant Nos. 11874059 and 12174405); Zhejiang Provincial Natural Science Foundation (Grant No. LR19A040002) and Beijing National Laboratory for Condensed Matter Physics.

## Keywords

van der Waals heterostructure, multiferroic materials, topological magnetism, anomalous valley Hall effect.

Supplementary material for

"Ferroelectricity controlled chiral spin textures and anomalous valley Hall effect in the Janus magnet-based multiferroic heterostructure"

Yingmei Zhu[1], Qirui Cui[1,3], Jinghua Liang[1], Yonglong Ga[1], Hongxin Yang[1,2,*]

[1] *Ningbo Institute of Materials Technology and Engineering, Chinese Academy of Sciences, Ningbo 315201, China*

[2] *Center of Materials Science and Optoelectronics Engineering, University of Chinese Academy of Sciences, Beijing 100049, China*

[3] *Faculty of Science and Engineering, University of Nottingham Ningbo China, Ningbo 315100, China*

*Email: hongxin.yang@nimte.ac.cn

**Section S1: Phonon calculation**

The dynamical stability of the two-dimensional Janus La*XY* monolayer is checked using the finite displacement method. A 4×4×1 and 5×5×1 supercell and a 5×5×1 *k*-point grid is adopted to compute the force constants. The force constant matrices and phonon dispersion are calculated implementing by the PHONOPY code [1]. The calculated phonon band structures of La*XY* are shown in Figure S1. For LaClBr and LaBrI, it is clearly seen that no imaginary frequency is present, indicating that the systems are dynamically stable. For LaClI, the acoustic phonon modes display negative frequencies around the Γ-point, which is related to in-plane bending of the two different halogen atomic planes [2,3].

**Section S2: Molecular dynamics simulation**

To verify the thermodynamic stability of La*XY* monolayer, we perform *ab initio* molecular dynamics (AIMD) simulations in the canonical *NVT* ensemble with Nosé thermostat [4]. A 4×4×1 supercell and a 5×5×1 *k*-point mesh is used. The simulations are conducted for 10 ps at 500K, and each time step is set to 1 fs. Figure S2 shows the evolution of the total energy and temperature during the AIMD simulations. And the structures (Figure S2) after simulations for 10 ps are still maintained at initial phase.

**Section S3: La*XY*/In$_2$Se$_3$ heterostructure (HS) configurations**

From the previous research, each atomic layer in In$_2$Se$_3$ monolayer only one element, which the atoms arranged in an equilateral triangular lattice and initially places on one of the three sublattice sites, A, B, or C, as shown in Figure S3. We mainly study the ferroelectric modulation of the topological magnetic phase and electronic states of La*XY* layers, then fixing the lattice of La*XY* can avoid the influence of external strain degree. As the lattice of In$_2$Se$_3$ layer is fixed, for example, LaClBr layer is stretched 0.51% to match In$_2$Se$_3$. The calculated magnetic parameters of LaClBr monolayer under 1% tensile strain are similar with the pristine LaClBr (Table S2). The same approach is also adopted in the multiferroic van der Waals heterostructures. For example, electric polarization manipulates a variety of properties of the magnetic anisotropy of the CrGeTe$_3$ in CrGeTe$_3$/In$_2$Se$_3$ HS, the conductivity of the CrI$_3$ in the CrI$_3$/Sc$_2$CO$_2$ HS, and the magnetic order of bilayer CrI$_3$ in the bilayer CrI$_3$/Sc$_2$CO$_2$ HS [5-7]. Considering structural symmetries of In$_2$Se$_3$ and La*XY*, we propose 12 stacking configurations of La*XY*/In$_2$Se$_3$ HSs by shifting or rotating the position of La*XY* layer. The site of each atom in La*XY* is provided at the top of each stacking sequence. The total energies of 12 optimized structures for La*XY*/P↓, and the energy for La*XY*/P↑ corresponding to the most stable stacking configuration is labeled in Figure S4. We find In$_2$Se$_3$ prefers to contact the heavier elements-terminated side in the *X* and *Y* elements. The most stable stacking sequence of LaClBr/In$_2$Se$_3$ is configuration BAB', which La and *X*/*Y* atoms of La*XY* sitting at the top-Se and bottom-In sites of In$_2$Se$_3$. For LaClI (LaBrI)/In$_2$Se$_3$, the configuration BCB', with La and *X*/*Y* atoms of La*XY* sitting at the top and bottom-In sites of In$_2$Se$_3$, is the most stable. To clarify the influence of the interface contact between two layers, we calculate the magnetic parameters of LaClBr/In2Se3 with Br-Se and Cl-Se contact as an example (Table S3). When the interface contact changes from Br-Se to Cl-Se, the DMI chirality is reversed. The calculated layer-resolved SOC energy difference $\Delta E_{soc}$ of Br-Se and Cl-Se contacts (see Figure S5) indicate that the DMI mainly origin from Br layer and the sign of $\Delta E_{soc}$ is opposite at different cases, which responds for the inverse chirality of DMI.

**Section S4: Magnetic anisotropy energy *K*, exchange interaction constant *J*, and Dzyaloshinskii-Moriya interaction $d_\parallel$ of La*XY* monolayer and La*XY*/P heterostructures**

The calculated methods of magnetic parameters *J*, *K*, and $d_\parallel$ are as follows: (i) The magnetic anisotropy energy *K* is defined as the energy difference between in-plane [100] and out-of-plane [001] magnetizations, $E = E_{100} - E_{001}$. The layer-resolved MAE is the integration of MAE of atoms in each layer. (ii) In order to determine the nearest-neighboring exchange coupling *J*, we compare the energy difference of ferromagnetic and antiferromagnetic states in a 2×1×1 supercell, which is expressed as $J = (E_{AFM} - E_{FM})/8$. When we calculate the NN exchange coupling for magnetization along positive directions of *x*, *y*, and *z* axis ($J_{xx}$, $J_{yy}$, and $J_{zz}$), the SOC effects are considered. Furthermore, for obtaining the second-nearest- and third-nearest-neighboring exchange coupling $J_2$ and $J_3$ of the systems, the total energies of the four spin configurations shown in Figure S5 are obtained as: $E_{FM} = -24J_1 - 24J_2 - 24J_3 + E_{other}$, $E_{AFM1} = 8J_1 + 8J_2 - 24J_3 + E_{other}$, $E_{AFM2} = 8J_1 - 8J_2 + 8J_3 + E_{other}$, $E_{AFM3} = -8J_1 + 8J_2 + 8J_3 + E_{other}$. For LaClBr/P↓ HS, the calculated $J_1$, $J_2$ and $J_3$ is 8.341, -0.427 and -0.012 meV. We find the magnitude of $J_1$ is much larger than that of $J_2$ and $J_3$, and is very close to the that directly obtained from the 2×1 supercell. Then we only consider the nearest-neighboring exchange coupling in the following calculation. (iii) In calculations of the in-plane component of DMI strength $d_\parallel$ [8], a 4×1×1 supercell and a Γ-centered 6×24×1 k-point mesh is adopted. We calculate the chirality-dependent energy difference (CDED) between the energies clockwise and anticlockwise chirality [see Figure 1] to obtain $d_\parallel$, $d_\parallel = (E_{cw}-E_{acw})/12$. To prove the effect of different supercells on DMI, we calculate DMI strength, $d_\parallel = \frac{E_{cw}-E_{acw}}{3nsin(2\pi/n)}$ ($n$=3,4) [8], by constructing 3×1×1 and 4×1×1 supercell using LaClBr/P↓ as an example. The calculated $d_\parallel$ is 0.192 and 0.197 meV for 3×1×1 and 4×1×1 supercells of LaClBr/P↓, respectively. It indicates that different supercells have little influence on DMI strength. By comparing the atom-layer energy between clockwise and anticlockwise configurations with SOC effects. In addition, we also consider the additional, symmetric, off-diagonal contributions to the exchange tensor, which is expressed by $\Gamma_{xy}$, $\Gamma_{xz}$, and $\Gamma_{yz}$ in the exchange matrix [9]. For example, to determine the $\Gamma_{xy}$ of LaClBr/In$_2$Se$_3$, we use a 2×1×1 supercell and compare the energies of Mn atoms with {($S$ 0 0), (0 $S$ 0)} and {(-$S$ 0 0), (0 $S$ 0)} spin configurations. The corresponding energy is $E_1 = J_{xx} + J_{yy} + 4\Gamma_{xy}$ and $E_2 = J_{xx} + J_{yy} - 4\Gamma_{xy}$. Accordingly, $\Gamma_{xy} = (E_1 - E_2)/8$. $\Gamma_{xy}$, $\Gamma_{xz}$, and $\Gamma_{yz}$ of LaClBr/P↓ is -0.048, 0.081, and 0.236 μeV, respectively. Compared with DMI components, the magnitude of Γ is very limited. These results indicate that Γ is negligible for La*XY*/In$_2$Se$_3$ systems. In order to demonstrate the validity of results obtained from CDED approach, we perform the four state method where we consider a 4×4×1 supercell of LaClBr monolayer with four spin configurations of La pairs as **S**$_1$={(S,0,0),(0,0,S)}, **S**$_2$={(S,0,0),(0,0,-S)}, **S**$_3$={(-S,0,0),(0,0,S)}, and **S**$_4$={(-S,0,0),(0,0,-S)}. In these four states, the spins of all the other spin sites are the same and are along the *y* axis, and corresponding $d_\parallel$ equals $\frac{-E_1-E_4+E_2+E_3}{4S^2}$, where the $d_\parallel$>0 and $d_\parallel$<0 favor spin configuration with anticlockwise and clockwise, respectively [10]. The calculated DMI strength of LaClBr using four state method and CDED approach is -0.015 and -0.012 meV, respectively, and both favor spin configuration with clockwise chirality. We further calculate the basic properties of LaClBr/P↓ with different cutoff energies (see Table S1). One can see that the results of *K*, *J*, and $d_\parallel$ is very similar when the different cutoff energy is adopted. Notably, all calculations are proceeded under the plane-wave basis set in 420 eV.

**Section S5: Atomistic spin model simulation-VAMPIRE**

All atomistic spin model simulations of La$XY$/In$_2$Se$_3$ HSs are performed using the VAMPIRE package [11]. For topological spin textures of these heterostructures, a 300 nm×300 nm square system is created with periodic boundary. To describe the dynamic of atomic spins, the Landau-Lifshitz-Gilbert (LLG) equation is adopted. The LLG can be expressed as: $\frac{\partial \boldsymbol{S}_i}{\partial t} = -\frac{\gamma}{(1+\lambda^2)}[\boldsymbol{S}_i \times \boldsymbol{B}_{eff}^i + \lambda \boldsymbol{S}_i \times (\boldsymbol{S}_i \times \boldsymbol{B}_{eff}^i)]$, where $\boldsymbol{S}_i$ is a unit vector representing the direction of the magnetic spin moment of site $i$, $\gamma$ is the gyromagnetic ratio, and the damping constant $\lambda$ is set to 0.2. $\boldsymbol{B}_{eff}^i$ is the net magnetic field on each spin and is obtained from $\boldsymbol{B}_{eff}^i = -\frac{1}{\mu_S}\frac{\partial H}{\partial \boldsymbol{S}_i}$, where $H$ is spin Hamiltonian of La$XY$/In$_2$Se$_3$ HSs.

Table S1. Calculated magnetic anisotropy energy $K$ (meV), nearest-neighbor exchange coupling $J$ (meV), and in-plane DMI component $d_{\parallel}$ (meV) of LaClBr/P↓ with different cutoff energy.

| cutoff energy | $K$ | $J_{xx}$ | $J_{yy}$ | $J_{zz}$ | $J$ | $d_{\parallel}$ |
|---|---|---|---|---|---|---|
| 420 eV | 0.010 | 7.809 | 7.803 | 7.788 | 7.800 | 0.197 |
| 520 eV | 0.010 | 7.837 | 7.830 | 7.816 | 7.828 | 0.197 |

Table S2. Table R1. Calculated lattice constants $a$ (Å), magnetic moment of La atom $m_{La}$ ($\mu_B$), magnetic anisotropy energy $K$ (meV), nearest-neighbor exchange coupling $J$ (meV), and in-plane DMI component $d_{\parallel}$ (meV) of La$XY$ monolayer, LaClBr monolayer under hole-doping, tensile (1%) and compressive (-1%) strain, and La$XY$/In$_2$Se$_3$ ($X$, $Y$ = Cl, Br, I).

|  |  | $a$ | $m_{La}$ | $K$ | $J_{xx}$ | $J_{yy}$ | $J_{zz}$ | $J$ | $d_{\parallel}$ |
|---|---|---|---|---|---|---|---|---|---|
| LaClBr | pristine | 4.087 | 0.385 | -0.010 | 6.518 | 6.515 | 6.502 | 6.512 | -0.012 |
|  | hole-doping | 4.087 | 0.373 | 0.014 | 7.406 | 7.400 | 7.386 | 7.397 | 0.146 |
|  | 1% | 4.128 | 0.382 | -0.011 | 6.423 | 6.419 | 6.398 | 6.413 | -0.014 |
|  | -1% | 4.046 | 0.388 | 0.032 | 6.565 | 6.562 | 6.553 | 6.560 | -0.010 |
| LaClI | - | 4.184 | 0.374 | -0.140 | 6.164 | 6.170 | 6.106 | 6.147 | -0.051 |
| LaBrI | - | 4.235 | 0.371 | -0.223 | 6.088 | 6.089 | 5.998 | 6.058 | -0.068 |
| LaClBr/P↓ | - | 4.087 | 0.368 | 0.010 | 7.809 | 7.803 | 7.788 | 7.800 | 0.197 |
| LaClBr/P↑ | - | 4.087 | 0.381 | 0.009 | 6.757 | 6.754 | 6.739 | 6.750 | -0.008 |
| LaClI/P↓ | - | 4.184 | 0.351 | -0.050 | 8.141 | 8.124 | 8.073 | 8.113 | 0.597 |
| LaClI/P↑ | - | 4.184 | 0.364 | -0.090 | 7.071 | 7.069 | 7.013 | 7.051 | 0.167 |
| LaBrI/P↓ | - | 4.235 | 0.343 | -0.011 | 8.280 | 8.271 | 8.239 | 8.263 | 0.424 |
| LaBrI/P↑ | - | 4.235 | 0.358 | -0.073 | 7.281 | 7.277 | 7.222 | 7.260 | 0.158 |

Table S3. Calculated magnetic moment of La atom $m_{La}$ ($\mu_B$), equilibrium interlayer distance $d$ (Å), magnetic anisotropy energy $K$ (meV), nearest-neighbor exchange coupling $J$ (meV), and in-plane DMI component $d_\parallel$ (meV) of LaClBr/P↓(↑) with Br-Se and Cl-Se contact.

| contact | system | $m_{La}$ | $d$ | $K$ | $J_{xx}$ | $J_{yy}$ | $J_{zz}$ | $J$ | $d_\parallel$ |
|---|---|---|---|---|---|---|---|---|---|
| Br-Se | LaClBr/P↓ | 0.368 | 3.073 | 0.010 | 7.809 | 7.803 | 7.788 | 7.800 | 0.197 |
| | LaClBr/P↑ | 0.381 | 3.101 | 0.009 | 6.757 | 6.754 | 6.739 | 6.750 | -0.008 |
| Cl-Se | LaClBr/P↓ | 0.365 | 2.952 | 0.005 | 7.842 | 7.836 | 7.819 | 7.832 | -0.130 |
| | LaClBr/P↑ | 0.381 | 2.987 | 0.008 | 6.747 | 6.744 | 6.731 | 6.741 | 0.331 |

Table S4. The in-plane DMI component $d_\parallel$ of La$XY$/In$_2$Se$_3$ heterostructures with P↓, P↑ and centrosymmetric configurations.

|  | P↓ | P↑ | Centrosymmetric |
|---|---|---|---|
| LaClBr/In$_2$Se$_3$ | 0.197 | -0.008 | 0.127 |
| LaClI/In$_2$Se$_3$ | 0.597 | 0.167 | 0.406 |
| LaBrI/In$_2$Se$_3$ | 0.424 | 0.158 | 0.285 |

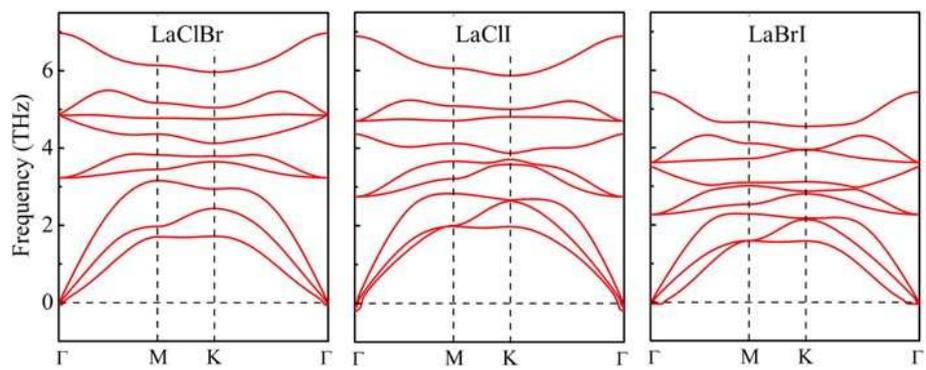

Figure S1. Calculated phonon band structure for the LaClBr, LaClI, and LaBrI monolayer.

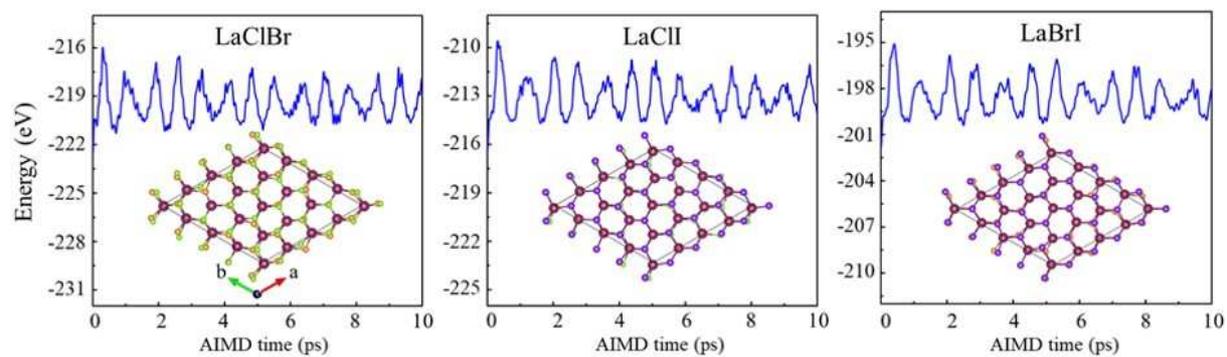

Figure S2. The evolution of the total internal energy and temperature during ab initio MD simulations for LaClBr, LaClI, and LaBrI monolayer at 500K.

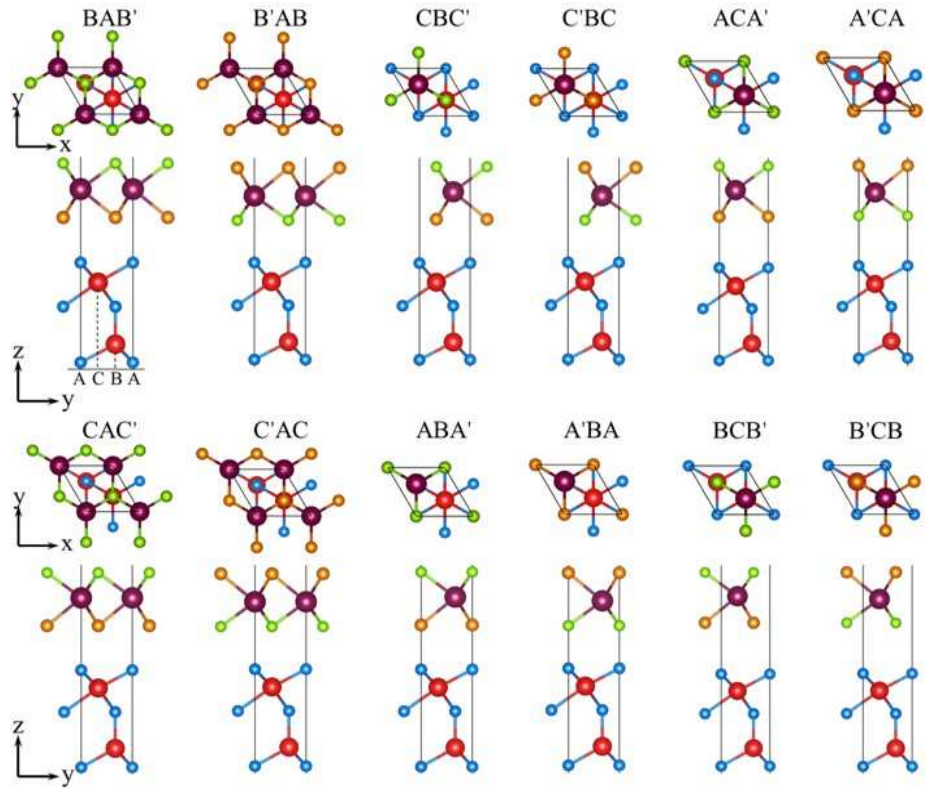

Figure S3. Top and side views of 12 stacking configurations for La$XY$/In$_2$Se$_3$ heterostructures, which is constructed by shifting La$XY$ monolayer along [1$\bar{1}$0] and rotating 60° around La. The site of each atom in La$XY$ is provided at the top of each stacking sequence.

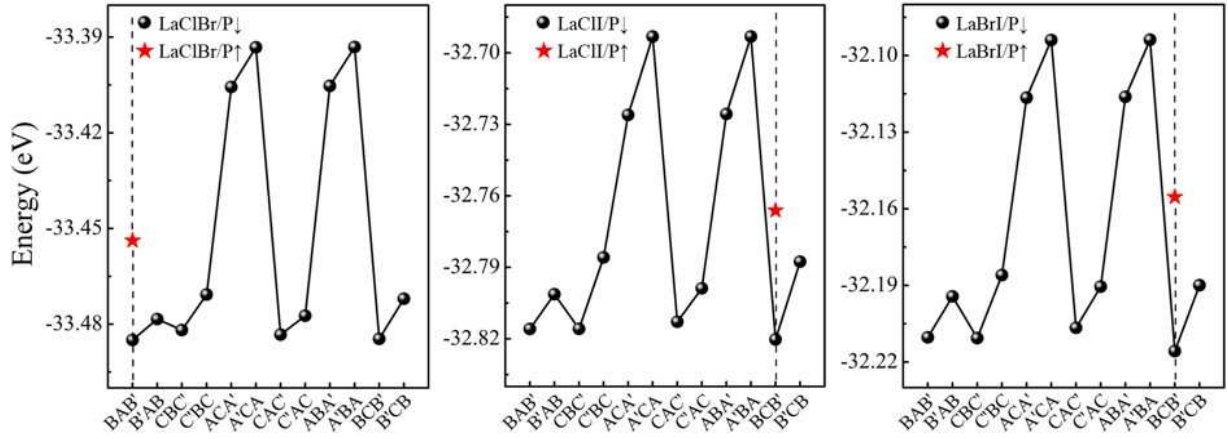

Figure S4. Total energies of 12 stacking configurations for La*XY*/In$_2$Se$_3$(P↓) heterostructures. The dashed line indicates the most stable stacking and the star represents the total energy of the corresponding La*XY*/In$_2$Se$_3$(P↑) structure.

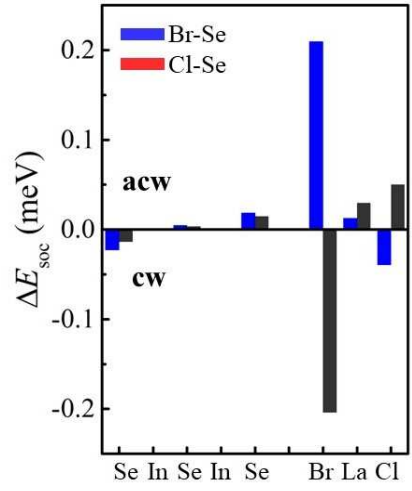

Figure S5. Layer-resolved localization of SOC energy difference $\Delta E_{soc}$ of LaClBr/P↓ with Br-Se and Cl-Se interface contacts. The positive(negative) values correspond for the contribution to DMI with anticlockwise (clockwise) chirality.

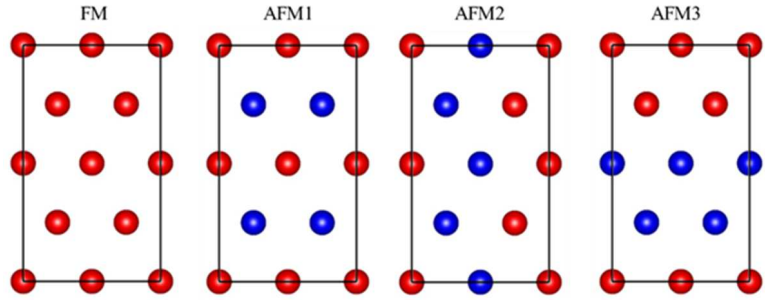

Figure S6. Four different magnetic configurations applied to calculate the second-nearest- and third-nearest-neighboring exchange coupling parameters. Red and blue balls represent the spin-up and spin-down of La atoms, respectively.

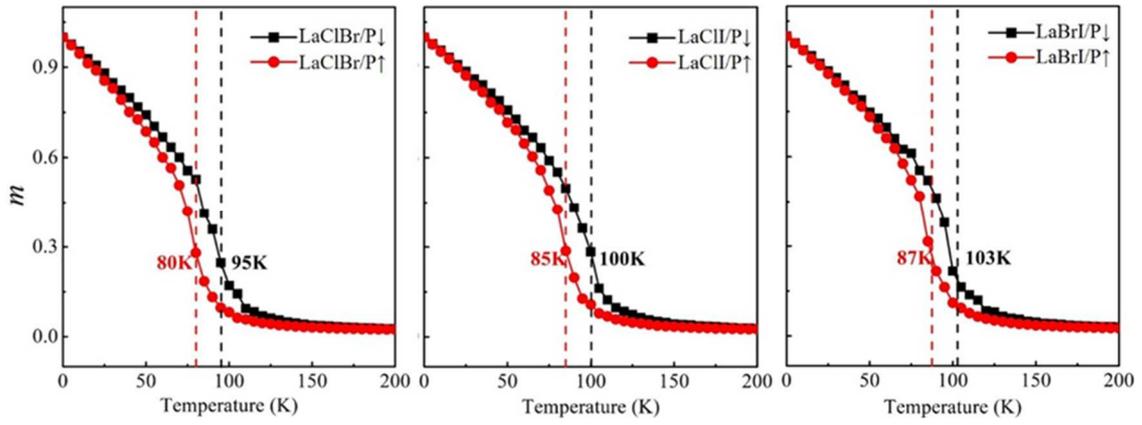

Figure S7. The Curie temperature ($T_c$) of LaXY/P↓(↑) heterostructures.

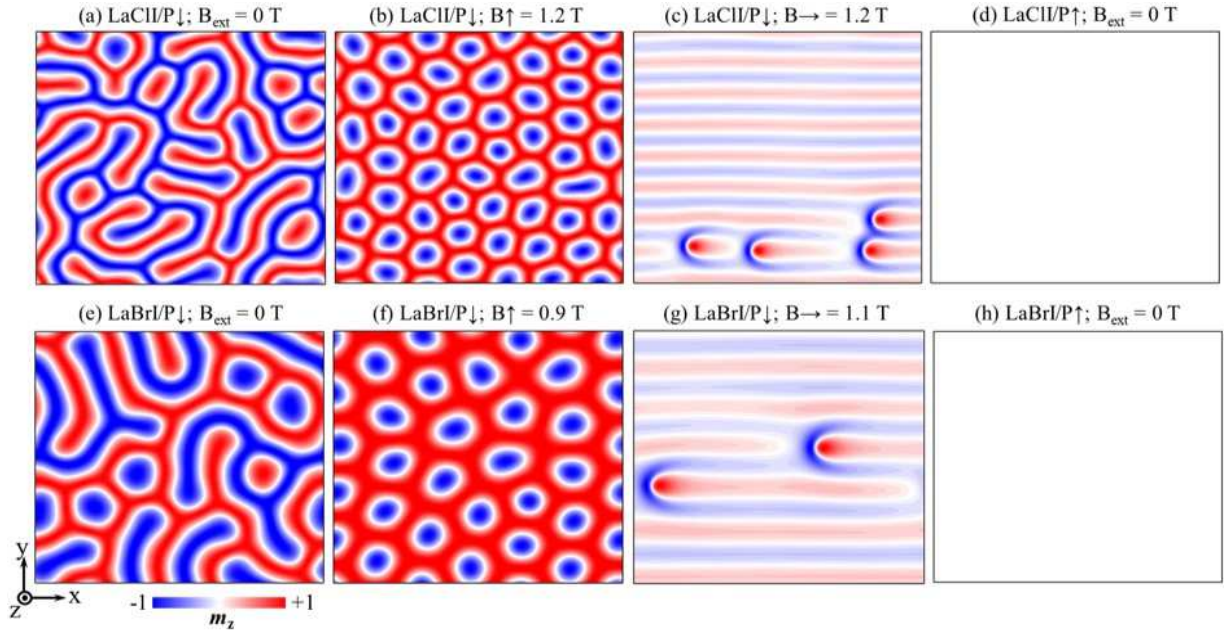

Figure S8. The spin textures for (a) LaClBr/P↓, (d) LaClBr/P↑, (e) LaBrI/P↓, and (h) LaBrI/P↑ in real space. (b), (f) Out-of-plane and (c), (g) in-plane external field of La$XY$/P↓ are simulated. The color map indicates the out-of-plane spin component of La atoms.

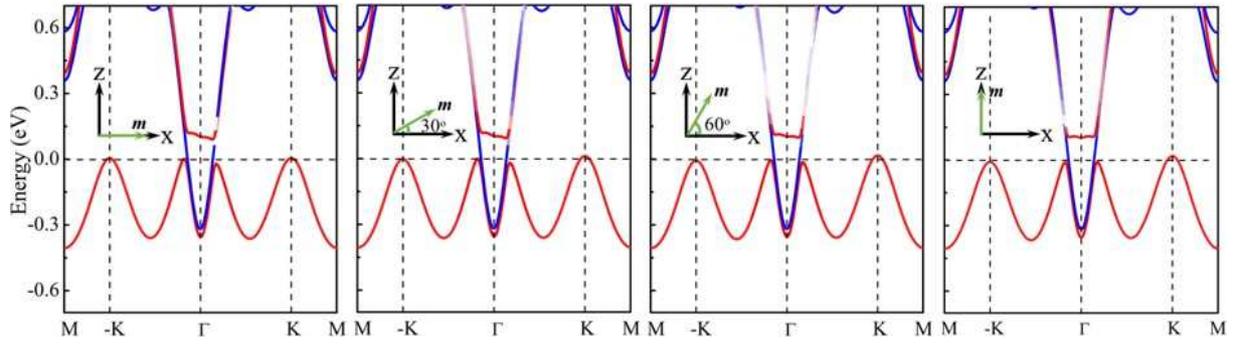

Figure S9. The LaClBr/In$_2$Se$_3$ heterostructure maintains metal states when the orientation of magnetization rotates from [100] to [001].

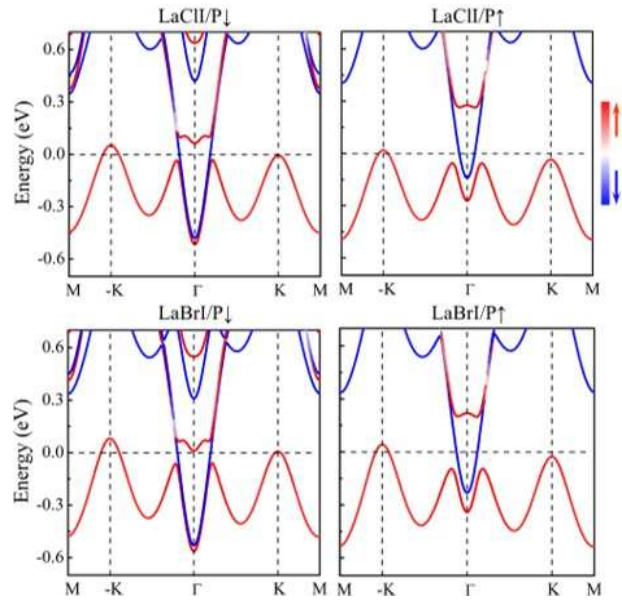

Figure S10. Spin-resolved band structure of (a) LaClI/P↓, (b) LaClI/P↑, (c) LaBrI/P↓, (b) LaBrI/P↑.

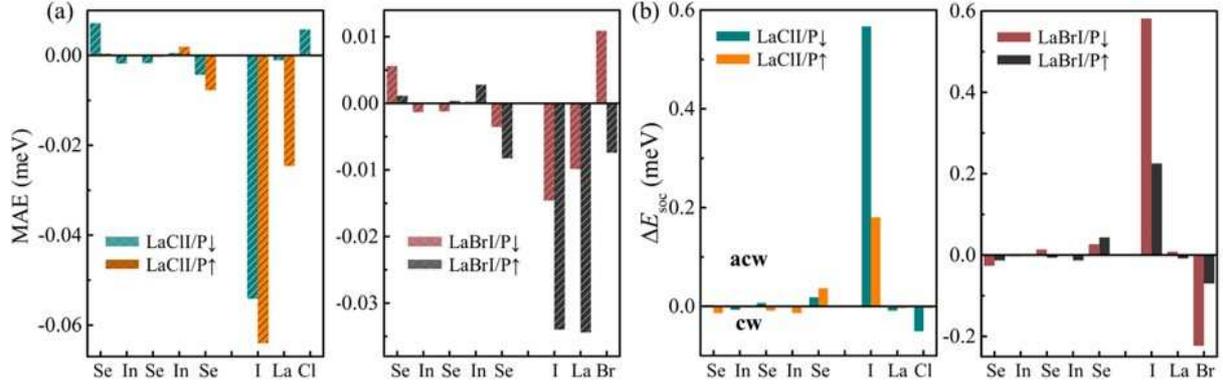

Figure S11. (a) Layer-resolved MAE of LaClI/P↓ (green shaded bar), LaClI/P↑ (orange shaded bar), LaBrI/P↓ (brown shaded bar), and LaBrI/P↑ (blue shaded bar). (b) Layer-resolved localization of the associated SOC energy difference $\Delta E_{SOC}$ between clockwise and anticlockwise spin configurations in LaClI/P↓ (green bar), LaClI/P↑ (orange bar), LaBrI/P↓ (brown bar), and LaBrI/P↑ (blue bar).

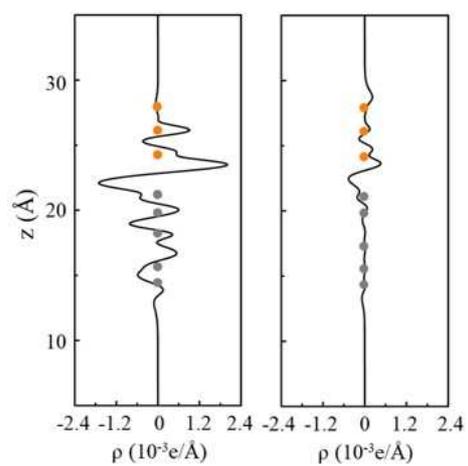

Figure S12. Planar-average charge-density difference of LaClBr/P↓ and LaClBr/P↑. The orange and grey dots represent the position of LaClBr and $In_2Se_3$ atoms, respectively.